# NEW INSIGHTS INTO THE NATURE OF THE ECLIPSING SYSTEM V609 AQUILAE

*By D. G. Turner[1,2]*
*Saint Mary's University, Halifax, Nova Scotia, Canada*

*E. A. Panko[3] and O. Sergienko*
*Kalinenkov Astronomical Observatory, Mykolaiv State University,*
*24 Nikolska str., Mykolaiv, Ukraine*

*D. J. Lane and D. J. Majaess[1]*
*Saint Mary's University, Halifax, Nova Scotia, Canada*

[1] Visiting Guest Investigator, Dominion Astrophysical Observatory, Herzberg Institute of Astrophysics, National Research Council of Canada.
[2] Visiting Astronomer, Harvard College Observatory Photographic Plate Stacks.
[3] Visiting Astronomer, Department of Astronomy and Physics, Saint Mary's University, Halifax, Nova Scotia, Canada.

A photometric study of the Near Contact Binary (NCB) system V609 Aql reveals it to be the westernmost star of a close double, with brightness variations and implied parameters more extreme than those derived in an earlier photographic study, in which images of the variable and companion were almost certainly blended. The system's brightness variations exhibit deep primary eclipses ($\Delta V = 1.04$) and secondary eclipses ($\Delta V = 0.44$) matched to a model fit with a derived orbital inclination of $i = 84°.8 \pm 0°.2$ and estimated component spectral types of F8-F9 and K2-K3. The primary overfills its Roche lobe in the optimum eclipse solution, inconsistent with the definition of NCBs. Period changes in the system are studied from 23 published times of light minimum and 21 newly-established values: 18 from examination of archival Harvard plates, and 3 from ASAS data and new CCD observations. O–C variations from 1891 to 2007 exhibit a long-term parabolic trend indicative of a period decrease, $dP/dt = -(7.75 \pm 1.39) \times 10^{-8}$ d yr$^{-1}$, corresponding to mass transfer to the secondary of $(6.5 \pm 1.2) \times 10^{-8}$ $M_\odot$ yr$^{-1}$. Superposed variations may indicate fluctuations in the mass flow. The system is estimated to be ~513 pc distant.

*Introduction*

The number of well-studied eclipsing binary systems in the Galaxy is a small fraction of the number listed in the 1969 and 1985 editions of the *General*

*Catalogue of Variable Stars*[1,2], and some systems with several bibliographic entries in the ADS (Astronomical Data System) may be relatively poorly studied. A good example is V609 Aquilae (2000 co-ordinates $20^h\ 09^m\ 58^s.77$, $+14°\ 38'\ 14".7$), originally described as an Algol-like variable[3], but now recognized as a Beta Lyrae system[4], primarily from a photographic study by Ishtchenko & Leibowitch[5]. The same study appears to be the basis for later summaries of more detailed system parameters and the object's designation as a detached system[6], as well as the cited spectral types of the component stars as F8 and G0[7]. With a period $P = 0^d.796565 = 19.1$ hours, V609 Aql undergoes a primary or secondary eclipse most clear evenings, making it ideal for observational study.

In 1994 V609 Aql was included in a List of Near Contact Binaries (NCB)[8], a new subclass of close binary systems (CBS) defined by Shaw[9]. NCB systems have periods of less than a day, exhibit the effects of tidal interaction, and have facing surfaces less than 0.1 orbital radius apart, but are not in contact. Such systems may be the evolutionary precursors to the A-type W UMa systems, and are probably in the early stages of mass transfer. NCBs may also be X-ray sources, although V609 Aql itself is not detected in the ROSAT All-Sky Survey[10].

In "An Atlas of O–C Diagrams of Eclipsing Binary Stars"[11] V609 Aql is noted to be a Beta Lyrae-type or W UMa-type system, but insufficiently researched. Only 19 times of light minimum over the interval 1937 to 1996 are cited, including unpublished results. Four additional times of light minimum since 2000 have been published[12,13,14,15], increasing the number of such estimates to 23. But the only publicly available light curve for the star[5] was derived from visual inspection of plates in the collection of Tashkent Astronomical Observatory, and is not of high precision.

Our interest in V609 Aql originated with its interesting O–C diagram, which displays irregular trends[11]. We therefore initiated a new study of the system in order to find additional times of light minimum, the intent being to clarify the nature of its period variability. V609 Aql was therefore observed photometrically to establish times for current minima, and was also investigated using the Harvard College Observatory Photographic Plate Collection to obtain archival or "historical" times of minima. As we describe here, our new observations for V609 Aql reveal it to be a much more interesting and different system than found by previous investigators.

*Observational Data and O–C Analysis*

New *V*-band observations were obtained for V609 Aql on 21 nights between 22 September and 18 December 2006 using a Celestron 28-cm Schmidt-Cassegrain telescope at the Abbey Ridge Observatory (Lane), an automated facility located at a dark site outside of Halifax, Nova Scotia. The telescope is equipped with a SBIG ST9 CCD camera and an Optec IFW filter wheel, although for the present study of V609 Aql only *V* observations were measured. All data represent means for combinations of short-exposure images that were analyzed by aperture photometry and normalized relative to GSC 01085-01422, our adopted reference star for the field (at an adopted magnitude of $V$ = 10.59), with HDE 354987 and three other stars in the field serving as check stars (see Table I and Fig. 1). The standard deviations for all observations of the reference star and check stars were ±$0^m.006$ to ±$0^m.008$, typical of other observations being made from the observatory. The data are summarized in Table II.

Additional observations of V609 Aql were obtained on one night (7 October 2006) using the 70-cm telescope of the Kalinenkov Astronomical Observatory of Nikolaev State University, equipped with a SBIG ST-7 camera in an instrumental photometric system closely approximating the standard $V$-band. The CCD exposures for the observations were of shorter duration than those obtained from Abbey Ridge, in an attempt to prevent image overlap for the newly-discovered companion star (see below). The resulting data represent means of 10 separate exposures, obtained by aperture photometry and tied to the same reference stars. Unfortunately, residual image overlap contaminated the photometry, making it difficult to calibrate relative to our reference stars, although the data display excellent overall agreement with our primary observations when combined with them after normalization (Table III).

All data for V609 Aql were phased using an existing ephemeris[11], namely:

$$HJD_{min} = 2429365.7284 + 0.7965639\ E,$$

where $E$ is the number of elapsed cycles.

The phased Abbey Ridge $V$-band observations of V609 Aql are plotted in Fig. 2. They were initially constrained to only a few hours on each clear night, but the star was also monitored continuously on three nights near the end of the run in an attempt to delineate critical portions of the light curve. Although full phase coverage for the binary system was not obtained, the data are sufficiently complete that one can map almost a complete light curve by using the symmetry of the light curves of close binaries and mirroring the data about primary minimum (lower portion of Fig. 2). Mirroring the light curve also allowed us to estimate the time of primary minimum accurately using the robust software used for matching Cepheid light curves[17], the result being a derived phase shift of $\Delta\varphi$ = −0.0097 ±0.0018, corresponding to O–C = +$0^d.0077$ ±0.0018, the uncertainty established by the fitting procedure used to minimize the resulting scatter in the data matched through mirroring. The corresponding time for primary minimum associated with the start of cycle 30989, following which we began a continuous run on V609 Aql, is HJD 2454050.4548 ±0.0014. That value is listed in Table IV.

In similar fashion the single-night observations for V609 Aql observed from Nikolaev were used to delineate secondary minimum for the system, and yielded an O−C datum through mirroring. That result is listed in the second-last row of Table IV.

We also obtained an estimate for the time of light minimum using $V$-data from the ASAS-3 (All Sky Automated Survey 3) project[17] matched to our resulting light curve adjusted for the phase offset. The ASAS-3 data exhibit a phase shift of $\Delta\varphi$ = −0.0432 relative to our best-fitting light curve, and also match the present observations to within a few millimagnitudes in *V*. The resulting O−C value (−0$^d$.0344) is listed in the third-last row of Table IV.

Additional estimates for times of light minimum in V609 Aql, both primary and secondary minima, were obtained through visual scanning of plates in the Harvard College Observatory Photographic Plate Collection, using suitable reference stars in the field for comparison[5]. A selection of patrol series plates was initially scanned for that purpose, but with limited success. Most patrol series exposures are roughly an hour in duration, which corresponds to a range in phase of ~0.05. By chance most early patrol series exposures of the field of V609 Aql correspond to times outside eclipse, and only ~1% of those examined (2) reveal the star near light minimum. Many higher resolution plates were exposed near times of mid-eclipse, however, and proved to be a more reliable source of data. Most of the high resolution plates have exposure times of order 10 minutes, corresponding to an uncertainty in phase of less than ±0.01.

In that manner we were able to estimate 18 times of light minimum for V609 Aql from 1891 to 1950, two being times of secondary minimum. The data are tabulated for reference purposes in Table IV.

The resulting O−C data are plotted in Fig. 3 (upper panel) as filled and open circles. From an analysis of all times of light minima we find the following parabolic solution:

$$\text{HJD}_{min} = (2429365.7233 \pm 0.0062) + (0.796566 \pm 0.000003) \, E \\ - (0.845 \pm 0.151) \times 10^{-10} \, E^2$$

The parabolic fit is shown in Fig. 3 (upper panel). The period is clearly decreasing, the measured rate being $dP/dt = -(7.75 \pm 1.39) \times 10^{-8}$ d yr$^{-1}$. There may be additional trends in the O−C variations, but the evidence is not well established.

The scatter resulting from meshing the mirrored observations with the actual observations is only about ±0$^m$.011 for the optimum fit, not much larger than the estimated uncertainties in the data. Given the range of nights and phase over which the observations were obtained, such small residuals suggest that any non-symmetric deviations of the actual light curve arising from starspots or an accretion disk are negligibly small. In fact, both features were considered in modeling the system, but were found to be unimportant.

*Light Curve Analysis*

It seems clear from our new light curve for V609 Aql (Fig. 2) that the system undergoes deeper eclipses than implied by the original photographic study[5]. Another unexpected discovery from the CCD observations is that the star is an optical double. The lower portion of Fig. 1 contains two CCD images, one near light minimum (lower left) and one near light maximum (lower right), illustrating that the eclipsing system is the westernmost (right hand) star of the pair. The two stars are only 5.6 arcseconds apart (Table I) and the pair is always blended in our observations, so the data of Table II refer to the combined light of both stars, as established from aperture photometry.

Reliable separation of the light of the variable from the combined light of the pair requires knowledge of the brightness of the contaminating star, which is not readily established. Our Nikolaev observations generated magnitude estimates for both stars separately, but contain residual contamination arising from scattered light from the neighbouring star, which was eliminated by normalizing the data to the Abbey Ridge data. We were also unable to derive reliable estimates for the two stars through crude profile fitting methods, owing to the pixel scale for the CCD. Instead, we were able to solve for the brightness of the (assumed) non-variable optical companion of V609 Aql by proceeding as follows.

Inspection of our image of the system near primary minimum indicates that the eclipsing system is marginally fainter than the companion at that epoch. The limits for eye detection of brightness differences between stars is typically $\sim0^m.1$[18], so we estimated the brightness of the optical companion by subtracting its light from the combined light data using various trial values, until the brightness of V609 Aql at primary minimum was $\sim0^m.1$ fainter. The result was a companion at $V = 12.35$, identical in brightness to check star C3, which lies conveniently close to the pair. The images of star C3 in the lower portion of Fig. 1 are sufficiently close in apparent size and brightness to those for the eastern optical companion of V609 Aql to confirm its inferred magnitude, although it is clearly a result in need of verification.

We summarize in Table V the inferred uncontaminated brightness of V609 Aql from the Abbey Ridge and Nikolaev observations, including mirrored values, and plot the data in Fig. 4. The results are clearly at variance with previous conclusions concerning the depths of primary and secondary minimum, and also about the brightness of the star in and out of eclipse (see Table VI).

We modeled the light curve for V609 Aql using *Binary Maker 3*[19] along with reasonable estimates for the properties of the two components. A low dispersion (120 Å mm$^{-1}$) spectrogram of the system near light maximum implies a spectral type of ~F8-9 V, consistent with previous estimates[7], but the spectral type of the fainter star in the system cannot be established directly, except through analysis of the light curve. We found through trial and error, in conjunction with examination of the residuals for various trials, the solution given in Table VI. An initial temperature for the primary was established from its F8-F9 V spectral type, corresponding to an intrinsic $B-V$ colour of 0.54-0.57, in conjunction with the $B-V$ colour-$T_{eff}$ relation of Gray[20], which implies $T_{eff}$ = 5988-6083 K. The final solution implies $T_{1eff}$ = 6050K and $T_{2eff}$ = 5000K, the latter corresponding to a K2-K3 dwarf. Both stars are larger than expected for dwarfs, presumably a consequence of stellar evolution and the changes occurring in a close binary system with mass transfer.

The mass ratio for eclipsing systems is difficult to establish without radial velocity information[21], although our solution yielded reasonable results. A mass of $M_1$ = 1.05 $M_\odot$ was adopted for the primary from its main-sequence spectral type F8-F9 and luminosity[22], and the eclipse solution yielded a reasonably well-defined mass ratio of $M_2/M_1 = 0.70 \pm 0.02$. The resulting implied secondary mass of $M_2 = 0.72 \pm 0.02\ M_\odot$ is, in fact, the value expected for a K2-K3 dwarf[22], apparently confirming the eclipse solution. Although we cannot exclude the possibility of systematic effects in our iterative technique, since the various system properties depend directly upon each other, the results should encourage others to observe the star in more comprehensive fashion to establish the system parameters more reliably. Precision radial velocity observations of the system are essential for confirming the inferred masses of the two components, as well as for constraining the light curve solution for the system, which is illustrated in the lower portion of Fig. 4.

A model for the V609 Aql system from the eclipse solution is given in Fig. 5. The primary star in the system is indicated to overfill its Roche lobe, since an alternate solution for a star just filling its Roche surface produced clear discrepancies with the observed light curve. The primary is therefore in the process of transferring mass to the secondary. The derived rate of period decrease can be used to find the rate of mass flow from the primary in the system in the case of conservative mass transfer[23], namely from:

$$\frac{dM_1}{dt} = \frac{1}{3P}\frac{dP}{dt}\left(\frac{M_1 M_2}{M_1 - M_2}\right).$$

In the present case the derived parameters for the system correspond to a rate of mass transfer to the secondary of $(6.5 \pm 1.2) \times 10^{-8}\ M_\odot$ year$^{-1}$, atypically large for a close binary system but not for a system in which one star overfills its Roche lobe. The fact that matter from the primary in the system is flowing directly to the secondary without the presence of an accretion disk presumably accounts for the lack of X-rays from the system.

The light curve of V609 Aql bears some similarity to the light curves of the Near Contact Binaries WZ Cyg[24] and MT Her[25], which have eclipse depths and inferred parameters like those of V609 Aql. However, neither of those systems displays the extreme light curve curvature of V609 Aql, and both are undergoing period increases rather than a period decrease. They are likely in a different phase of evolution as close binary systems than V609 Aql. In fact, given all of the available evidence, V609 Aql no longer satisfies the criteria of a Near Contact Binary (NCB), since the primary overfills its Roche surface and both components are oversized. In contrast, most NCBs and their subclasses consist of a primary at or near its Roche lobe, with only the secondary in the system being oversized[8].

*The Distance to V609 Aql*

The eclipse solution for V609 Aql allows one to establish a reasonably reliable estimate for the distance to the system. Although the components are non-spheroidal and in near-contact, the view of the system during secondary eclipse is toward the side of the primary facing away from the secondary. One can adjust the orbital inclination in *Binary Maker 3* so that secondary eclipse is an occultation, the effective radius of the primary at that instant being 1.846 $R_\odot$ for the component masses estimated previously. With the effective temperature estimated from the model (which we assume applies to the back hemisphere) and inferred parameters for the Sun[26], we obtain a luminosity for the back hemisphere of the primary of 4.095 $L_\odot$, or $M_V$ = +3.29. The light originating from the primary during secondary eclipse is $V$ = 11.84, resulting in an observed distance modulus of $V-M_V$ = 8.55.

It can be argued that V609 Aql must be unreddened. The available broadband *B* and *V* magnitudes, primarily photographic, refer to the combined light of the optical double, whereas the *JHK* colours[27] for the individual stars imply that the eastern component is a red star, with $J-H$ colour comparable to unreddened late-type stars of $B-V \cong +1.1$. The $J-H$ colour for V609 Aql is that of a yellow dwarf, but uncertainties of as much as ±0.1 in the *JHK* colours make it difficult to be more specific.

A more direct estimate is made as follows. The assumed spectral type of F8-F9 for the primary corresponds to $(B-V)_0 \cong +0.55$. Our photographic estimates of the brightness of V609 Aql on the Harvard plates, and the original study of Ishtchenko & Leibowitch[5], imply $B \cong 11.7$ for the combined light of the optical double at light maximum. The eastern component should have $B \cong 13.45$ according to our estimates for its visual brightness and broad band colour, making the brightness of the eclipsing component roughly $B \cong 11.94$ at maximum brightness. The observed visual maximum brightness is $V$ = 11.39, resulting in an observed colour of $B-V \cong +0.55$, identical to the expected unreddened broad band colour. Small changes to the calculations, within the magnitude of potential uncertainties in the estimates, do not alter the results significantly, so it seems clear that any interstellar reddening of the system must be negligibly small.

The field of V609 Aql at Galactic co-ordinates $l$ = 55°.20, $b$ = −10°.03 is adjacent to nearby Galactic fields where the reddening has been studied previously[28]. In those fields stars appear to be unreddened at distances of up to ~500 pc, beyond which dust extinction produces colour excesses of $E_{B-V}$ = 0.3 or larger. Since V609 Aql is unreddened, its implied distance modulus is $V_0-M_V$ = 8.55, corresponding to $d$ = 513 pc ($\pi$ = 0″.0019), *i.e.* at roughly the maximum distance beyond which dust extinction becomes important. The solution is consistent with the known spatial distribution of dust in adjacent Galactic fields[28], but is larger than the original estimate of the system's distance[6], which was 385 pc. The difference appears to originate in the larger radii found here for the primary and secondary stars in the system.

*Conclusions*

From 18 new times of light minimum for V609 Aql obtained from examination of images in the Harvard College Observatory Photographic Plate

Collection and three new estimates for light minimum obtained from recent observations of the system, we have calculated a new ephemeris for V 609 Aql that includes a parabolic term. The inferred rate of period decrease, $dP/dt = -(7.75 \pm 1.39) \times 10^{-8}$ d yr$^{-1}$, implies a rate of mass transfer of $(6.5 \pm 1.2) \times 10^{-8}$ $M_\odot$ year$^{-1}$ from the primary in the system to the less massive component, the primary overfilling its Roche lobe according to the new eclipse solution presented here. Small irregularities in the rate of period decrease for the system may simply be indicative of inhomogeneities in the rate of mass flow between the two components. The inferred parameters are at variance with the characteristics displayed by other members of the class of Near Contact Binaries[8], of which V609 Aql can no longer be considered a member.

TABLE I
*Reference Stars for Observations of V609 Aql*

| Star | RA(2000) | DEC(2000) | $V$ | Notes |
|---|---|---|---|---|
| V609 Aql  | $20^h\ 09^m\ 58^s.58$ | $+14°\ 38'\ 12''.9$ | var   | |
| Companion | $20^h\ 09^m\ 58^s.94$ | $+14°\ 38'\ 12''.9$ | 12.35 | See text, 5″.6 separation |
| Std       | $20^h\ 10^m\ 17^s.43$ | $+14°\ 35'\ 07''.1$ | 10.59 | GSC 01085-01422 |
| C1        | $20^h\ 10^m\ 21^s.40$ | $+14°\ 36'\ 28''.8$ | 9.77  | HDE 354987 |
| C2        | $20^h\ 10^m\ 16^s.74$ | $+14°\ 41'\ 01''.3$ | 11.07 | |
| C3        | $20^h\ 09^m\ 52^s.00$ | $+14°\ 38'\ 07''.9$ | 12.35 | |
| C4        | $20^h\ 10^m\ 03^s.65$ | $+14°\ 40'\ 26''.7$ | 12.68 | |

TABLE II
*New Observations for V609 Aql and companion*

| HJD | Phase | V | HJD | Phase | V |
|---|---|---|---|---|---|
| 2454000.5525 | 0.3628 | 11.095 | 2454050.5003 | 0.0668 | 11.292 |
| 2454000.5510 | 0.3608 | 11.095 | 2454050.4975 | 0.0633 | 11.292 |
| 2454000.6086 | 0.4332 | 11.229 | 2454050.5027 | 0.0699 | 11.265 |
| 2454000.6985 | 0.5461 | 11.230 | 2454050.5074 | 0.0758 | 11.242 |
| 2454001.6424 | 0.7309 | 11.026 | 2454050.5125 | 0.0821 | 11.221 |
| 2454001.6819 | 0.7806 | 11.047[a] | 2454050.5174 | 0.0882 | 11.195 |
| 2454005.5400 | 0.6240 | 11.082 | 2454050.5222 | 0.0943 | 11.179 |
| 2454005.5951 | 0.6932 | 11.036 | 2454050.5272 | 0.1006 | 11.172 |
| 2454006.5393 | 0.8785 | 11.115 | 2454050.5319 | 0.1064 | 11.156 |
| 2454006.6468 | 0.0134 | 11.555 | 2454050.5371 | 0.1130 | 11.137 |
| 2454009.5083 | 0.6058 | 11.107 | 2454050.5417 | 0.1188 | 11.125 |
| 2454009.5469 | 0.6542 | 11.057 | 2454050.5468 | 0.1251 | 11.111 |
| 2454009.6679 | 0.8061 | 11.066 | 2454050.5517 | 0.1313 | 11.111 |
| 2454014.5487 | 0.9334 | 11.320 | 2454050.5567 | 0.1377 | 11.099 |
| 2454014.5896 | 0.9848 | 11.605 | 2454050.5620 | 0.1443 | 11.093 |
| 2454015.5801 | 0.2283 | 11.034 | 2454050.5667 | 0.1502 | 11.090 |
| 2454015.6180 | 0.2758 | 11.018 | 2454050.5711 | 0.1556 | 11.076 |
| 2454016.5038 | 0.3879 | 11.120 | 2454051.4683 | 0.2821 | 11.028 |
| 2454016.5404 | 0.4338 | 11.218 | 2454051.5133 | 0.3385 | 11.057 |
| 2454017.6510 | 0.8280 | 11.069 | 2454061.4318 | 0.7902 | 11.035 |
| 2454019.5127 | 0.1652 | 11.053 | 2454061.4318 | 0.7902 | 11.035 |
| 2454024.4986 | 0.4245 | 11.178 | 2454061.4364 | 0.7958 | 11.040 |
| 2454024.5393 | 0.4756 | 11.304 | 2454061.4412 | 0.8019 | 11.044 |
| 2454025.4686 | 0.6422 | 11.066 | 2454061.4462 | 0.8082 | 11.054[a] |
| 2454025.5140 | 0.6991 | 11.027 | 2454061.4509 | 0.8141 | 11.058[a] |
| 2454025.6108 | 0.8207 | 11.066 | 2454061.4561 | 0.8206 | 11.047[a] |
| 2454038.5357 | 0.0465 | 11.384 | 2454061.4607 | 0.8264 | 11.060[a] |
| 2454038.5954 | 0.1214 | 11.133 | 2454061.4661 | 0.8331 | 11.058[a] |
| 2454041.4385 | 0.6907 | 11.028 | 2454061.4722 | 0.8409 | 11.080 |
| 2454041.4484 | 0.7031 | 11.031 | 2454061.4771 | 0.8470 | 11.083 |
| 2454041.4554 | 0.7119 | 11.029 | 2454061.4822 | 0.8534 | 11.095 |
| 2454041.4624 | 0.7207 | 11.020 | 2454061.4870 | 0.8594 | 11.111 |
| 2454041.4694 | 0.7295 | 11.027 | 2454061.4923 | 0.8661 | 11.111 |
| 2454041.4764 | 0.7383 | 11.014 | 2454061.4970 | 0.8720 | 11.120 |
| 2454041.4834 | 0.7470 | 11.022 | 2454061.5020 | 0.8783 | 11.148 |
| 2454041.4904 | 0.7558 | 11.020 | 2454061.5069 | 0.8844 | 11.151 |
| 2454041.4970 | 0.7641 | 11.032 | 2454061.5118 | 0.8906 | 11.165 |
| 2454041.5053 | 0.7745 | 11.021 | 2454061.5169 | 0.8970 | 11.190 |
| 2454041.5108 | 0.7815 | 11.028 | 2454061.5217 | 0.9029 | 11.207 |
| 2454050.4434 | 0.9954 | 11.641 | 2454061.5269 | 0.9095 | 11.230 |
| 2454050.4434 | 0.9954 | 11.641 | 2454061.5316 | 0.9154 | 11.245 |
| 2454050.4479 | 0.0010 | 11.623 | 2454061.5366 | 0.9217 | 11.296 |
| 2454050.4528 | 0.0071 | 11.609 | 2454061.5400 | 0.9259 | 11.298 |
| 2454050.4579 | 0.0136 | 11.570 | 2454062.5201 | 0.1563 | 11.082 |
| 2454050.4627 | 0.0196 | 11.536 | 2454081.4531 | 0.9247 | 11.284 |
| 2454050.4680 | 0.0262 | 11.498 | 2454082.4309 | 0.1522 | 11.084 |
| 2454050.4828 | 0.0448 | 11.398 | 2454082.4723 | 0.2042 | 11.037 |
| 2454050.4876 | 0.0509 | 11.357 | 2454088.4494 | 0.7077 | 11.018 |
| 2454050.4928 | 0.0573 | 11.324 | | | |

[a] Observation of low quality.

TABLE III
*Nikolaev Observations of V609 Aql*

| HJD | Phase | V |
|---|---|---|
| 2454029.2410 | 0.3781 | 11.538 |
| 2454029.2486 | 0.3876 | 11.565 |
| 2454029.2597 | 0.4015 | 11.613 |
| 2454029.2657 | 0.4091 | 11.626 |
| 2454029.2734 | 0.4187 | 11.693 |
| 2454029.2793 | 0.4262 | 11.674 |
| 2454029.2871 | 0.4360 | 11.712 |
| 2454029.2932 | 0.4436 | 11.766 |
| 2454029.3064 | 0.4602 | 11.795 |
| 2454029.3239 | 0.4821 | 11.842 |
| 2454029.3349 | 0.4959 | 11.830 |
| 2454029.3527 | 0.5182 | 11.756 |
| 2454029.3597 | 0.5270 | 11.708 |
| 2454029.3743 | 0.5454 | 11.675 |
| 2454029.3804 | 0.5530 | 11.657 |
| 2454029.3851 | 0.5589 | 11.626 |

TABLE IV
*Newly Derived Epochs of Minimum Light for V609 Aql*

| HJD$_{min}$(obs) | Cycle (*E*) | HJD$_{min}$(calc) | O−C (d) | Year |
|---|---|---|---|---|
| 2411964.6344 | −21845.0 | 2411964.7900 | −0.1556 | 1891 |
| 2412348.5912 | −21363.0 | 2412348.7338 | −0.1426 | 1892 |
| 2412360.5611 | −21348.0 | 2412360.6823 | −0.1212 | 1892 |
| 2412368.5377 | −21338.0 | 2412368.6479 | −0.1102 | 1892 |
| 2414925.5472 | −18128.0 | 2414925.6180 | −0.0708 | 1899 |
| 2414933.5536 | −18118.0 | 2414933.5837 | −0.0301 | 1899 |
| 2414947.4798 | −18100.5 | 2414947.5235 | −0.0437 | 1899 |
| 2415289.5874 | −17671.0 | 2415289.6477 | −0.0603 | 1900 |
| 2415548.9022 | −17345.5 | 2415548.9293 | −0.0271 | 1901 |
| 2415618.5976 | −17258.0 | 2415618.6286 | −0.0310 | 1901 |
| 2415724.5048 | −17125.0 | 2415724.5716 | −0.0668 | 1901 |
| 2416639.7569 | −15976.0 | 2416639.8235 | −0.0666 | 1904 |
| 2416785.5235 | −15793.0 | 2416785.5947 | −0.0712 | 1904 |
| 2417094.5966 | −15405.0 | 2417094.6615 | −0.0649 | 1905 |
| 2418543.5561 | −13586.0 | 2418543.6113 | −0.0552 | 1909 |
| 2421092.5566 | −10386.0 | 2421092.6157 | −0.0591 | 1916 |
| 2428426.5910 | −1179.0 | 2428426.5796 | +0.0114 | 1936 |
| 2433484.7483 | +5171.0 | 2433484.7603 | −0.0120 | 1950 |
| 2453872.7789 | +30766.0 | 2453872.8133 | −0.0344 | 2006 |
| 2454029.3323 | +30962.5 | 2454029.3382 | −0.0059 | 2006 |
| 2454050.4548 | +30989.0 | 2454050.4471 | +0.0077 | 2006 |

TABLE V
*Reduced Observations for V609 Aql*

| Phase | V | Phase | V | Phase | V | Phase | V |
|---|---|---|---|---|---|---|---|
| 0.0002 | 12.439 | 0.1811 | 11.452 | 0.5056 | 11.842 | 0.8254 | 11.436 |
| 0.0059 | 12.402 | 0.1869 | 11.446 | 0.5082 | 11.830 | 0.8255 | 11.463 |
| 0.0104 | 12.365 | 0.1891 | 11.463 | 0.5196 | 11.826 | 0.8300 | 11.445 |
| 0.0120 | 12.373 | 0.1932 | 11.432 | 0.5275 | 11.795 | 0.8313 | 11.455 |
| 0.0183 | 12.267 | 0.1993 | 11.426 | 0.5306 | 11.756 | 0.8329 | 11.468 |
| 0.0184 | 12.296 | 0.2050 | 11.419 | 0.5394 | 11.708 | 0.8380 | 11.452 |
| 0.0245 | 12.230 | 0.2050 | 11.419 | 0.5441 | 11.766 | 0.8389 | 11.486 |
| 0.0311 | 12.159 | 0.2091 | 11.422 | 0.5509 | 11.709 | 0.8395 | 11.478 |
| 0.0496 | 11.982 | 0.2137 | 11.409 | 0.5517 | 11.712 | 0.8430 | 11.489 |
| 0.0514 | 11.958 | 0.2146 | 11.436 | 0.5578 | 11.675 | 0.8450 | 11.498 |
| 0.0557 | 11.913 | 0.2207 | 11.399 | 0.5613 | 11.690 | 0.8457 | 11.484 |
| 0.0617 | 11.852 | 0.2310 | 11.415 | 0.5615 | 11.674 | 0.8509 | 11.502 |
| 0.0622 | 11.858 | 0.2331 | 11.418 | 0.5620 | 11.707 | 0.8518 | 11.488 |
| 0.0681 | 11.806 | 0.2393 | 11.398 | 0.5654 | 11.657 | 0.8575 | 11.511 |
| 0.0692 | 11.816 | 0.2481 | 11.400 | 0.5690 | 11.693 | 0.8583 | 11.505 |
| 0.0705 | 11.794 | 0.2569 | 11.389 | 0.5707 | 11.629 | 0.8639 | 11.529 |
| 0.0735 | 11.813 | 0.2642 | 11.406 | 0.5712 | 11.626 | 0.8643 | 11.529 |
| 0.0747 | 11.763 | 0.2657 | 11.408 | 0.5786 | 11.626 | 0.8700 | 11.529 |
| 0.0798 | 11.732 | 0.2745 | 11.398 | 0.5861 | 11.613 | 0.8709 | 11.529 |
| 0.0806 | 11.727 | 0.2806 | 11.395 | 0.6001 | 11.565 | 0.8737 | 11.561 |
| 0.0856 | 11.709 | 0.2833 | 11.410 | 0.6072 | 11.542 | 0.8763 | 11.549 |
| 0.0870 | 11.695 | 0.2869 | 11.409 | 0.6096 | 11.538 | 0.8768 | 11.542 |
| 0.0922 | 11.673 | 0.2874 | 11.395 | 0.6106 | 11.523 | 0.8822 | 11.567 |
| 0.0931 | 11.655 | 0.2921 | 11.413 | 0.6289 | 11.486 | 0.8831 | 11.584 |
| 0.0982 | 11.647 | 0.2960 | 11.408 | 0.6343 | 11.505 | 0.8833 | 11.535 |
| 0.0991 | 11.630 | 0.3020 | 11.420 | 0.6471 | 11.463 | 0.8887 | 11.596 |
| 0.1046 | 11.609 | 0.3045 | 11.409 | 0.6566 | 11.450 | 0.8892 | 11.588 |
| 0.1054 | 11.620 | 0.3410 | 11.450 | 0.6590 | 11.450 | 0.8946 | 11.620 |
| 0.1108 | 11.588 | 0.3434 | 11.450 | 0.6955 | 11.409 | 0.8954 | 11.609 |
| 0.1113 | 11.596 | 0.3529 | 11.463 | 0.6980 | 11.420 | 0.9009 | 11.630 |
| 0.1167 | 11.535 | 0.3657 | 11.505 | 0.7040 | 11.408 | 0.9018 | 11.647 |
| 0.1169 | 11.584 | 0.3711 | 11.486 | 0.7079 | 11.413 | 0.9069 | 11.655 |
| 0.1178 | 11.567 | 0.3894 | 11.523 | 0.7126 | 11.395 | 0.9078 | 11.673 |
| 0.1232 | 11.542 | 0.3904 | 11.538 | 0.7131 | 11.409 | 0.9130 | 11.695 |
| 0.1237 | 11.549 | 0.3928 | 11.542 | 0.7167 | 11.410 | 0.9144 | 11.709 |
| 0.1263 | 11.561 | 0.3999 | 11.565 | 0.7194 | 11.395 | 0.9194 | 11.727 |
| 0.1291 | 11.529 | 0.4139 | 11.613 | 0.7255 | 11.398 | 0.9202 | 11.732 |
| 0.1300 | 11.529 | 0.4214 | 11.626 | 0.7343 | 11.408 | 0.9253 | 11.763 |
| 0.1357 | 11.529 | 0.4288 | 11.626 | 0.7358 | 11.406 | 0.9265 | 11.813 |
| 0.1361 | 11.529 | 0.4293 | 11.629 | 0.7431 | 11.389 | 0.9295 | 11.794 |
| 0.1417 | 11.505 | 0.4310 | 11.693 | 0.7519 | 11.400 | 0.9308 | 11.816 |
| 0.1425 | 11.511 | 0.4346 | 11.657 | 0.7607 | 11.398 | 0.9319 | 11.806 |
| 0.1482 | 11.488 | 0.4380 | 11.707 | 0.7669 | 11.418 | 0.9378 | 11.858 |
| 0.1491 | 11.502 | 0.4385 | 11.674 | 0.7690 | 11.415 | 0.9383 | 11.852 |
| 0.1543 | 11.484 | 0.4387 | 11.690 | 0.7793 | 11.399 | 0.9443 | 11.913 |
| 0.1550 | 11.498 | 0.4422 | 11.675 | 0.7854 | 11.436 | 0.9486 | 11.958 |
| 0.1570 | 11.489 | 0.4483 | 11.712 | 0.7863 | 11.409 | 0.9504 | 11.982 |
| 0.1605 | 11.478 | 0.4491 | 11.709 | 0.7909 | 11.422 | 0.9689 | 12.159 |
| 0.1611 | 11.486 | 0.4559 | 11.766 | 0.7950 | 11.419 | 0.9755 | 12.230 |
| 0.1620 | 11.452 | 0.4606 | 11.708 | 0.7950 | 11.419 | 0.9816 | 12.296 |
| 0.1671 | 11.468 | 0.4694 | 11.756 | 0.8007 | 11.426 | 0.9817 | 12.267 |
| 0.1687 | 11.455 | 0.4725 | 11.795 | 0.8068 | 11.432 | 0.9880 | 12.373 |
| 0.1700 | 11.445 | 0.4804 | 11.826 | 0.8109 | 11.463 | 0.9896 | 12.365 |
| 0.1745 | 11.463 | 0.4918 | 11.830 | 0.8131 | 11.446 | 0.9941 | 12.402 |
| 0.1746 | 11.436 | 0.4944 | 11.842 | 0.8189 | 11.452 | 0.9998 | 12.439 |

TABLE VI
*Derived System Parameters for V609 Aql*

| Parameter | Ishtchenko & Leibowitch (1955) | Brancewicz & Dworak (1980) | This Paper |
|---|---|---|---|
| $V$ | … | … | 11.40 |
| $\Delta V_1$ | … | … | 1.04 |
| $\Delta V_2$ | … | … | 0.44 |
| $B$ | 11.7 | … | … |
| $\Delta B_1$ | 0.4 | … | … |
| $\Delta B_2$ | 0.2 | … | … |
| Separation | … | 4.97 $R_\odot$ | 4.39 $R_\odot$ |
| $R_1$ | … | 1.49 $R_\odot$ | 1.84 $R_\odot$ |
| $R_2$ | … | 1.24 $R_\odot$ | 1.47 $R_\odot$ |
| $RL_1$ | … | 74% | 113% |
| $RL_2$ | … | 71% | 98% |
| $L_1$ | … | 2.34 $L_\odot$ | 2.70 $L_\odot$ |
| $L_2$ | … | 1.43 $L_\odot$ | 0.80 $L_\odot$ |
| $T_1$ | … | 5870 K | 6050 ±25 K |
| $T_2$ | … | 5680 K | 5000 ±25 K |
| $M_1$ | … | 1.49 $M_\odot$ | 1.05 $M_\odot$ (adopted) |
| $M_2$ | … | 1.10 $M_\odot$ | 0.74 ±0.02 $M_\odot$ |
| $M_1/M_2$ | … | 0.74 | 0.70 ±0.02 |
| $Sp.T._1$ | … | F8 | F8-F9 |
| $Sp.T._2$ | … | … | K2-K3 |
| $i$ | … | … | 84°.8 ±0°.2 |

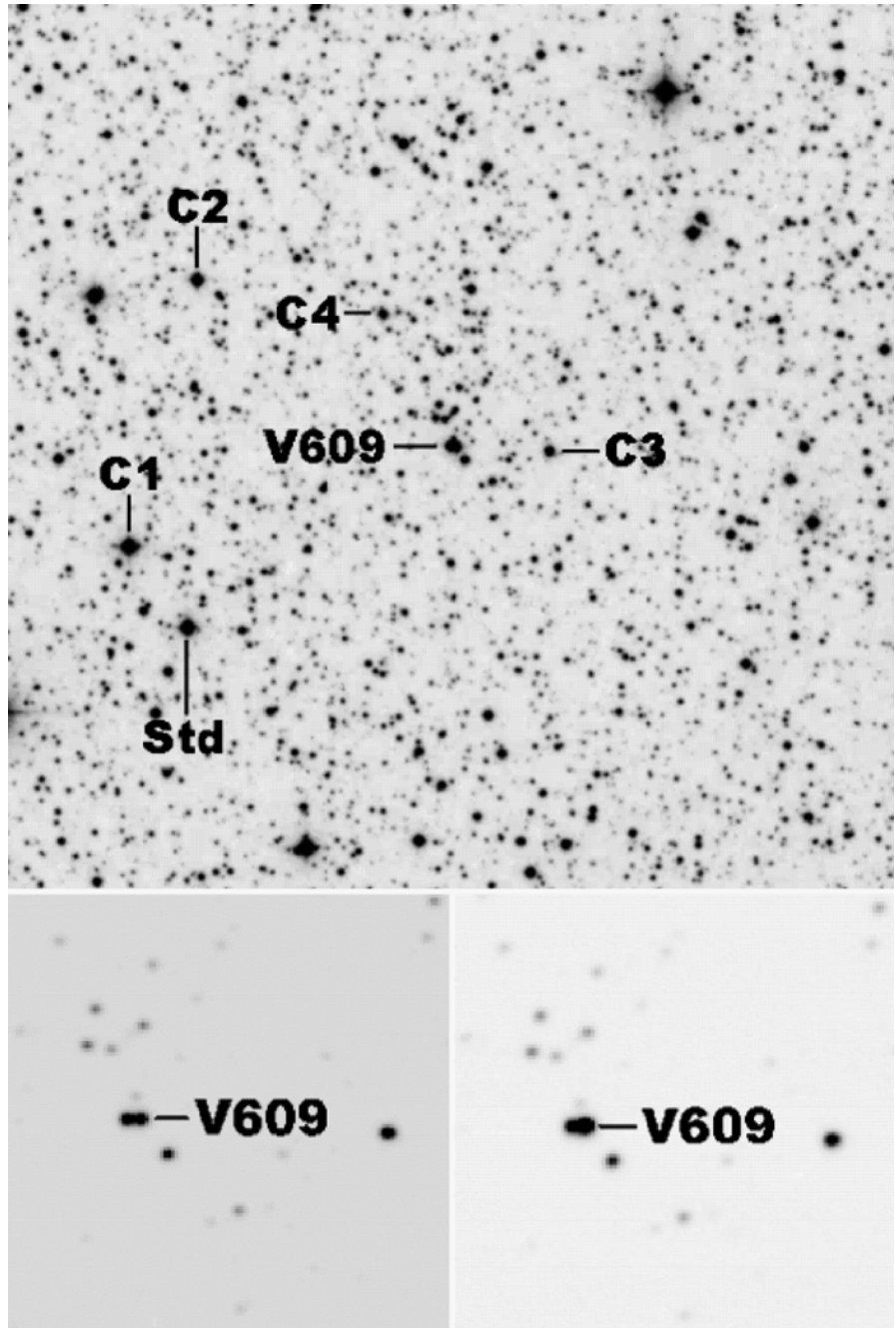

FIG. 1

A finder chart for the field centered on V609 Aql from the red image of the Palomar Observatory Sky Survey (top). The field of view measures 15′ × 15′ and shows the location of the variable, the adopted reference star, and four check stars used for the observations. The lower portion of the figure displays two enlargements from CCD images of V609 Aql at phases 0.9998 (lower left) and 0.8700 (lower right).

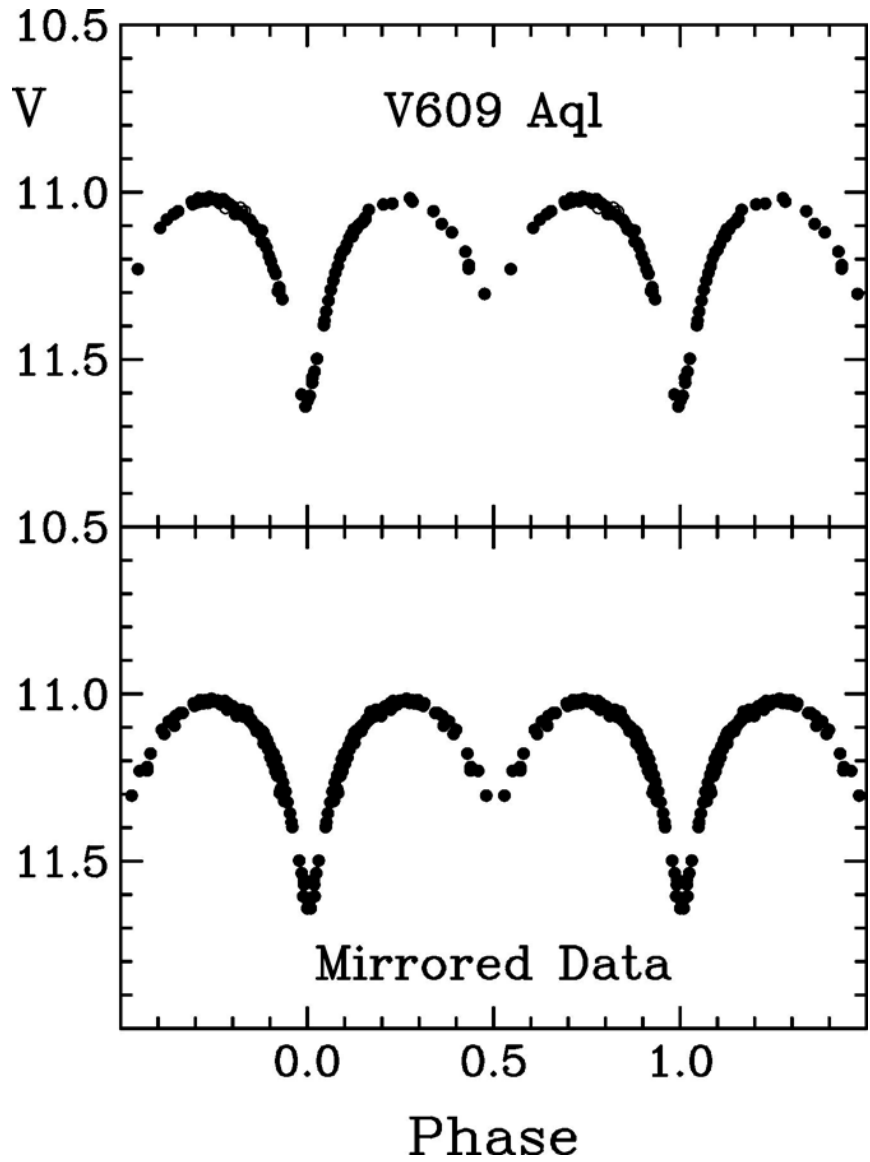

Fig. 2

*V*-band CCD observations for V609 Aql from the present program (upper). Open circles depict low quality observations. The lower portion of the figure displays the same data mirrored about zero phase.

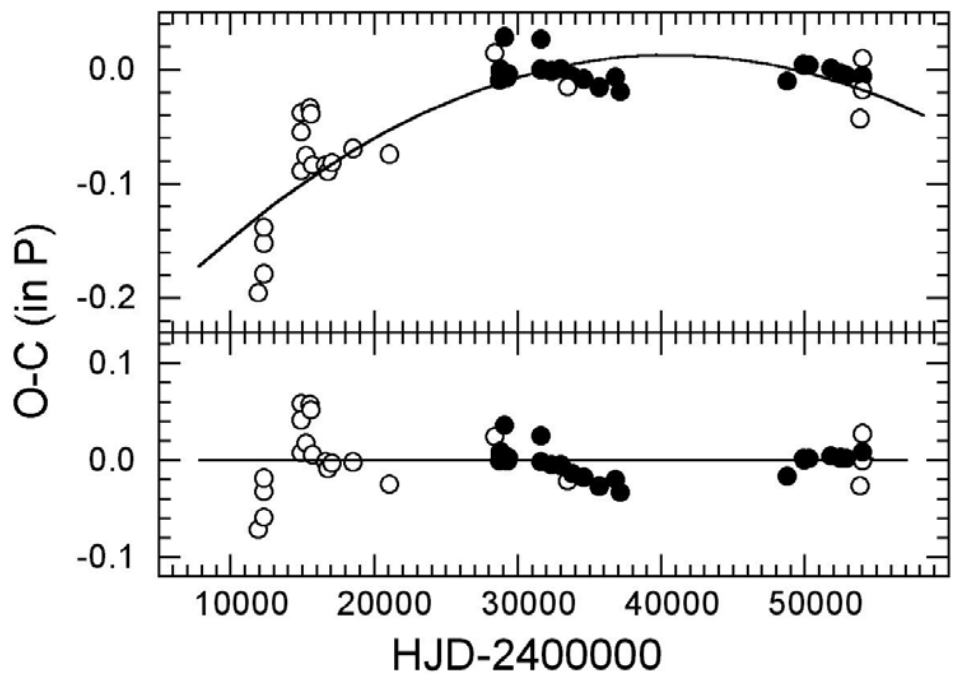

FIG. 3

O−C data (in units of phase offset) for V609 Aql plotted as a function of observed Heliocentric Julian Date of light minimum (upper). The lower plot shows the same data after removal of the parabolic trend evident in the data in the upper portion of the figure. Filled circles denote published times of light minimum, open circles the data of this paper.

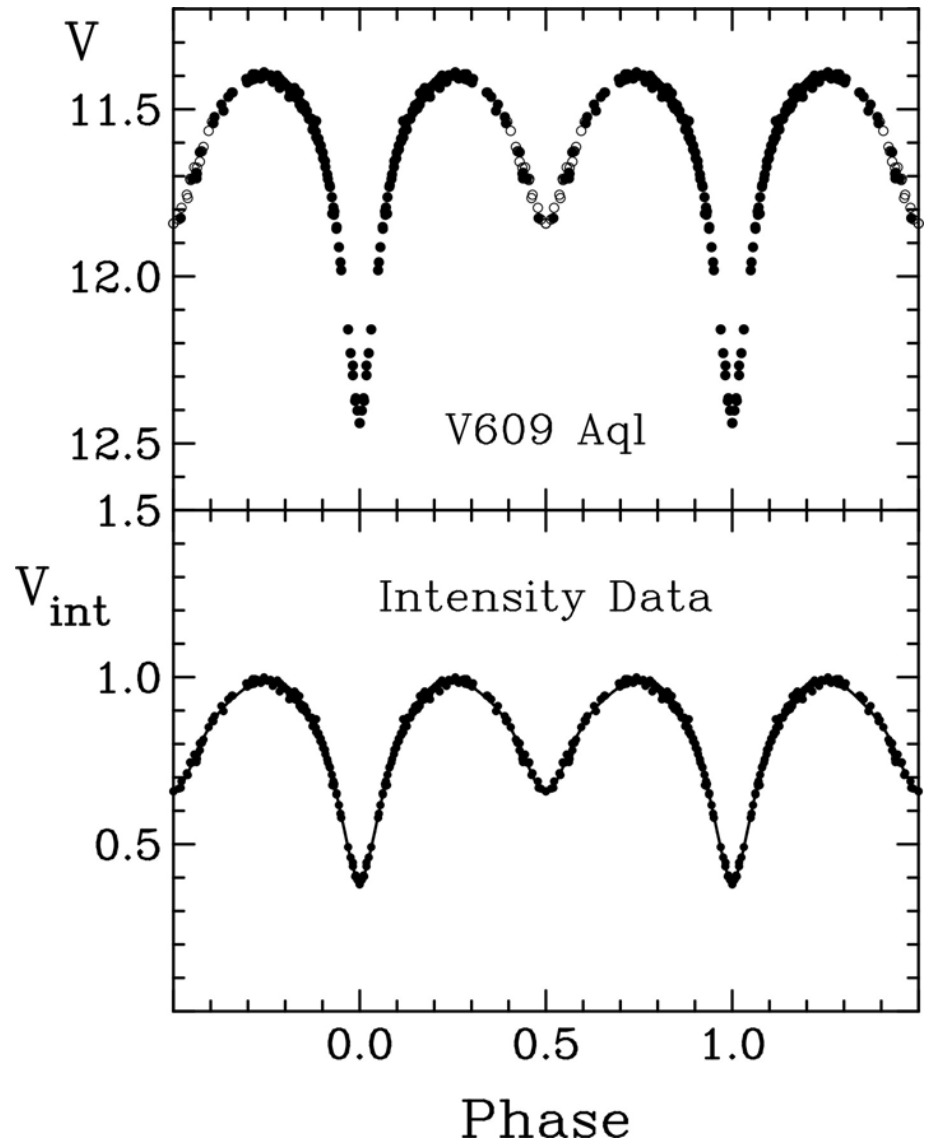

FIG. 4

*V*-band observations (including mirrored data) for V609 Aql (upper) corrected for contamination by a companion of $V = 12.35$. Open circles represent the Nikolaev observations normalized to the Abbey Ridge data. The same data are plotted as intensities (lower) along with the best-fitting model light curve.

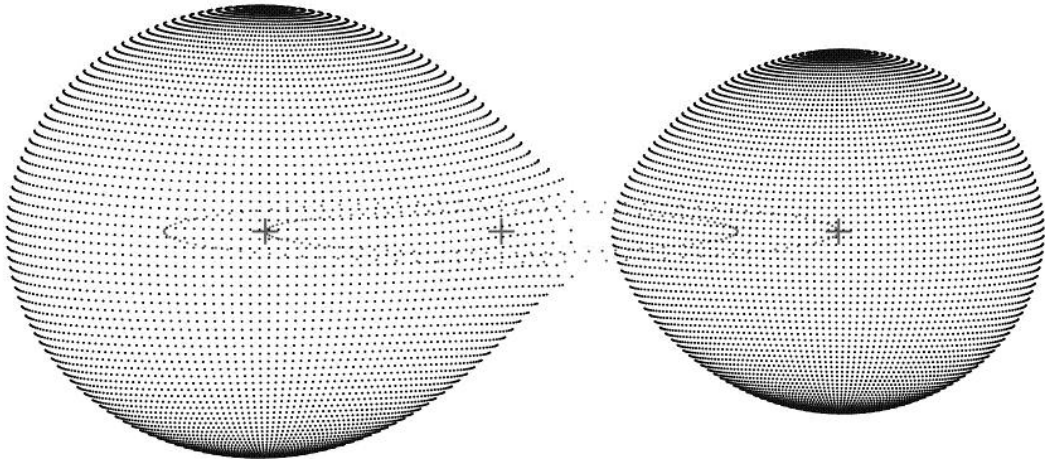

FIG. 5

A model for the V609 Aql system at phase 0.25 from *Binary Maker 3*.